\pdfoutput=1
\documentclass[letterpaper,conference,compsoc]{IEEEtran}
\usepackage[utf8]{inputenc}
\usepackage[T1]{fontenc}
\usepackage{microtype}
\ifCLASSOPTIONcompsoc
  \usepackage[nocompress]{cite}
\else
  \usepackage{cite}
\fi
\ifCLASSINFOpdf
  \usepackage[pdftex]{graphicx}
  \DeclareGraphicsExtensions{.pdf,.jpeg,.png}
\else
  \usepackage[dvips]{graphicx}
  \DeclareGraphicsExtensions{.eps}
\fi
\begin{document}
\title{Estimating Maximum Error Impact in Dynamic Data-driven Applications for Resource-aware Adaption of Software-based Fault-Tolerance}
\author{\IEEEauthorblockN{Björn Bönninghoff}
\IEEEauthorblockA{CS 12 DAES, TU Dortmund\\
Email: bjoern.boenninghoff@udo.edu}
\and
\IEEEauthorblockN{Horst Schirmeier}
\IEEEauthorblockA{CS 12 ESS, TU Dortmund\\
Email: horst.schirmeier@udo.edu}
}
\maketitle
\begin{abstract}
The rise of transient faults in modern hardware requires system designers to consider errors occurring at runtime. 
Both hardware- and software-based error handling must be deployed to meet application reliability requirements. 
The level of required reliability can vary for system components and depend on input and state, so that a selective use of resilience methods is advised, especially for resource-constrained platforms as found in embedded systems.
If an error occurring at runtime can be classified as having negligible or tolerable impact, less effort can be spent on correcting it.  
As the actual impact of an error is often dependent on the state of a system at time of occurrence, it can not be determined precisely for highly dynamic workloads in data-driven applications.
We present a concept to estimate error propagation in sets of tasks with variable data dependencies. 
This allows for a coarse grained analysis of the impact a failed task may have on the overall output. 
As an application example, we demonstrate our method for a typical dynamic embedded application, namely a decoder for the H.264 video format.
\end{abstract}
\IEEEpeerreviewmaketitle
\section{Introduction}
Due to reduced voltages and higher integration densities of integrated circuits, embedded systems hardware becomes more susceptible to transient faults caused by environmental influences like electromagnetic radiation or particle strikes\cite{irts2015}.
While a system %
must be protected against critical failure, 
some application domains may allow incorrect results to be produced, meaning that, e.g., Silent Data Corruption may be accepted. %
The impact an error has on the output, for example in the number of affected individual values, may vary depending on where and when it occurs. 
With resource restrictions typically found in embedded systems, an estimation of this impact becomes desirable to enable a selective application of fault-tolerance mechanisms. 
To provide this information, Engel, Schmoll et al.\cite{engel2011unreliable} propose a binary classification of individual data objects with respect to their resilience requirements, which is annotated for output data and propagated via type reasoning, enabling distinction between a possible system failure and mere quality degradation.
Hiller et al. present an elaborate model\cite{RDS2001:Jhumka} to analyze error-propagation between modules in distributed systems and provide a tool for offline analysis\cite{Hiller:2002:PEE:566171.566184}.
In this work, our goal is to provide a task-level error-propagation model that provides an upper bound for the possible error impact of a given task instance, that can be easily constructed for varying task dependencies as found in data-driven workloads. We then show the applicability of this model for an example of such data-driven application, a parallelized H.264 decoder.
\section{Error Propagation and Impact Estimation}
The objective of our model is to reflect error propagation
within a set of dependent tasks typically found in parallelized applications to give an estimation of maximum error impact regarding application output. 
\subsubsection*{Task and Communication Model}
We assume the following: A task has a given set of input channels that denote the data that has to be available for the task to start. Also, an outgoing channel denotes the output of a task. Each task will process its input and run to completion, where the output will become available both for reading by other task instances and/or as application output.
\subsubsection*{Error Propagation Graph}\label{SS:EPG}
From this model, we can construct an error propagation graph $EPG = \{V,E\}$, a directed graph composed by task-nodes $v \in V$ and propagation-edges $e \in {E}$.
Allowing multiple instances of the same task to be represented with individual nodes makes $EPG$ cycle-free. 
For all $v \in V$, we assume 
the worst case of any error being present in the input of a task resulting in erroneous output. 
For each node $v_i$, $m_{i,local}$ denotes the maximum error impact this node can locally have on the output.
To find an upper bound on the error impact, we need to determine $m_{i,global}$ for all nodes, which is the sum of $m_{i,local}$ and all $m_{j,local}$ where $v_j$ is a descendant of $v_i$.
\begin{figure}[th]
\centering
\includegraphics[width=8cm,height=3.5cm]{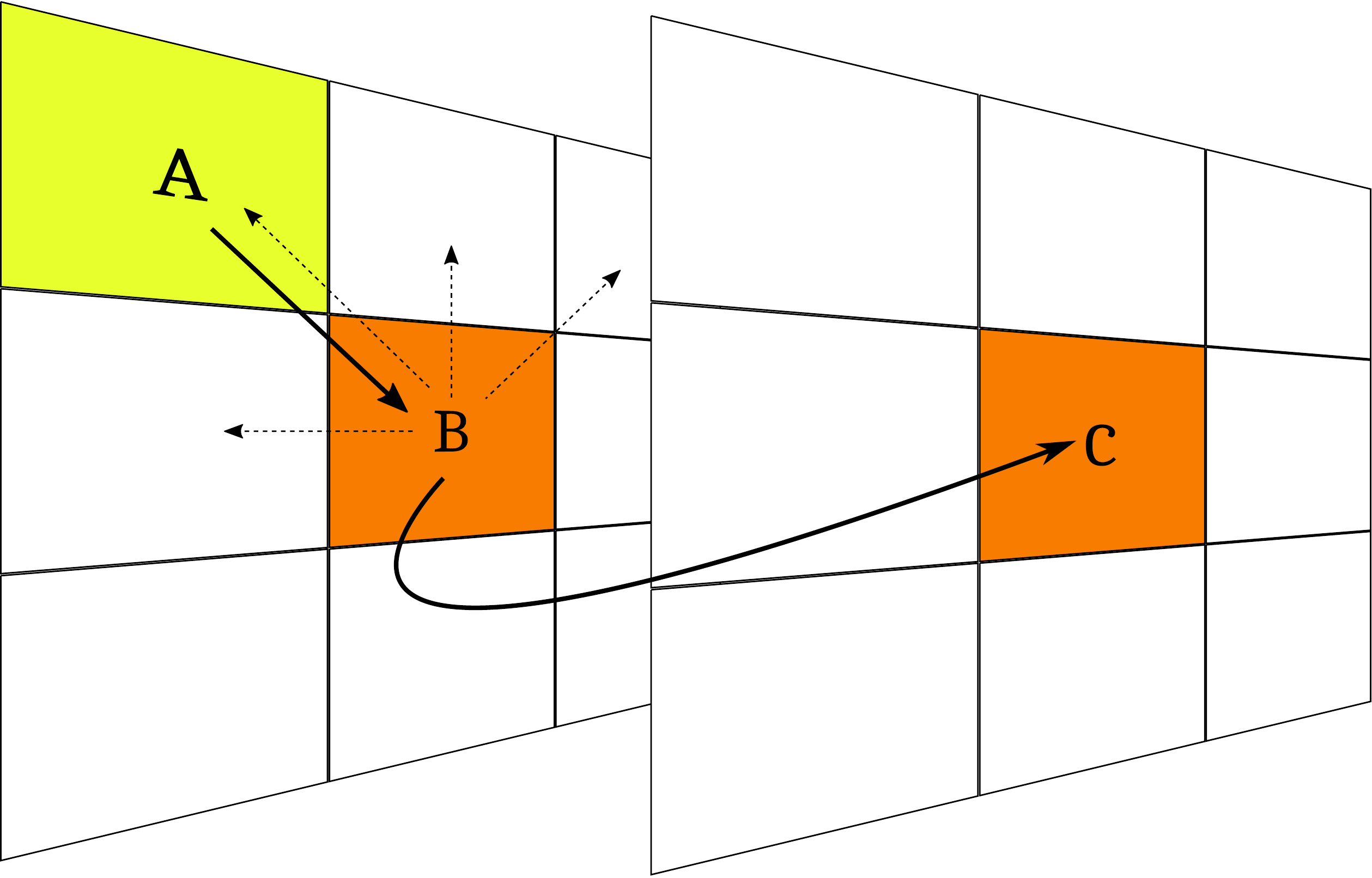}%
\caption{An error in the task constructing macroblock $A$ may propagate directly to $B$, and indirectly to macroblock $C$, which references $B$ via motion compensation. Dashed lines mark the possible references of intra-predicted macroblock B, solid lines denote propagation edges through actual references at runtime.}
\label{fig_ref}
\end{figure}
\section{Application Example}
To demonstrate a possible application of this method, we choose a H.264 video decoder application, with the features available for the baseline profile. 
We give a short introduction to the decoder structure and show that it exposes dynamic data-flow behavior,
and then present experimental results for an error-propagation analysis on exemplary input data. 
\subsubsection*{H.264 Decoder Structure}
The H.264 decoder iteratively constructs images from an incoming compressed datastream, containing control information as well as image information. 
Each image is constructed in macroblock units, square areas of 16x16 pixels in size, which are processed in a scanline manner. 
The input requires to be parsed in sequential manner, but parsing can be decoupled from macroblock reconstruction, which may then be performed in parallel tasks that follow a hybrid reconstruction approach\cite{Meenderinck2009}. 
In a first step, the image content is predicted from previously decoded data. 
In a second step, DCT-coefficients providing residual information undergo an inverse transformation and are added to the predicted image. 
The prediction can be distinguished into two methods, namely Intra-Prediction and Motion Compensation, both allowing for variable references to data from other macroblocks that allow errors to propagate.
\subsubsection*{Intra-Prediction}
Macroblocks using Intra-Prediction can be subdivided into up to 16 rectangular areas, each having a prediction mode. Depending on the selected mode, previously decoded image data from the surrounding area is sampled and extrapolated. As only available data can be accessed and is thus restricted by decoding order, only data from the topleft, top, topright and left neighboring block can be referenced (see Fig. \ref{fig_ref}). The actual subset of these four possible references can be derived directly from the available control information.
\subsubsection*{Motion Compensation}
In contrast to this, Motion Compensation is provided with a reference list to previously decoded frames. Again, the macroblock may be subdivided up to 16 times. Each area is provided with a reference index and corresponding motion vectors. Added to the current position of the area, they define an area in the reference frame from which the image data is used for prediction. As references are allowed on quarter-pixel accuracy, the referenced area also contains an additional margin for interpolation, resulting in the possiblity to cover up to 4 macroblocks.  
\subsubsection*{Macroblock-level Error Propagation}
\begin{figure}[t]
\includegraphics[]{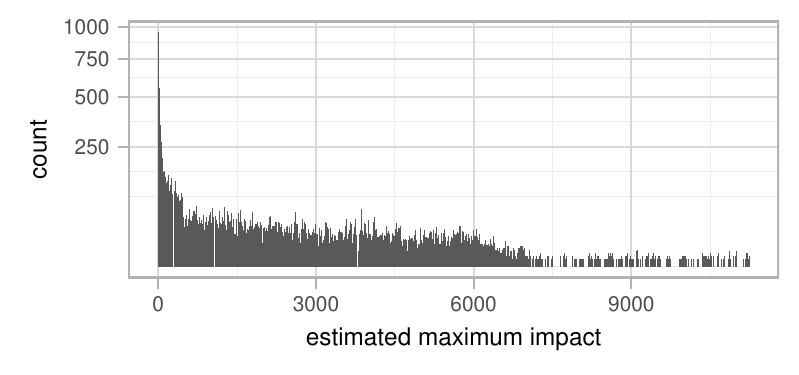}%
\caption{Histogram of estimated maximum impact from erronous macroblock output for a series of 150 frames from an example H.264 file (square-root scale on y-axis).}
\label{fig_sim}
\end{figure}

According to our model in Section \ref{SS:EPG}, we  construct an Error Propagation Graph while decoding the input bitstream. 
For each macroblock, we add a node $v_i$ to $V$, 
and normalize the maximum possible local impact $m_{i,local}$ to $1$. 
We use the macroblock reference information that becomes available to 
construct the propagation edges. 
For Intra-Prediction blocks, a maximum of $4$ edges are inserted into $E$, while motion-compensated macroblocks can introduce up to $16*4=64$ edges.
Once %
a %
graph is complete, meaning that no new references to its nodes can be constructed, which can be signaled in the input bitstream, we may then calculate the global impact $m_{i,global}$%
, which will then be available when starting the macroblock reconstruction tasks.
In Figure \ref{fig_sim}, we present a histogram of these values for a set of 150 consecutive frames, showing a wide distribution of estimated impact for individual macroblocks. 
\section{Conclusion}
We %
introduced our method to %
estimate maximum error impact in tasksets with data-driven dependencies. 
We showed the applicability to a real-world application, where we see large input-dependent variation of error impact for a given task. 
This information becomes available at runtime and can thus be used for a flexible, resource-aware deployment of SW fault-tolerance methods. 
\ifCLASSOPTIONcompsoc
  \section*{Acknowledgments}
\else
  \section*{Acknowledgment}
\fi
\noindent
This work was supported by the German Research Council (DFG) focus program SPP 1500 under grant MA 943/10-3.
\bibliographystyle{IEEEtran}
\bibliography{abs}

\begin{thebibliography}{1}
\providecommand{\url}[1]{#1}
\csname url@samestyle\endcsname
\providecommand{\newblock}{\relax}
\providecommand{\bibinfo}[2]{#2}
\providecommand{\BIBentrySTDinterwordspacing}{\spaceskip=0pt\relax}
\providecommand{\BIBentryALTinterwordstretchfactor}{4}
\providecommand{\BIBentryALTinterwordspacing}{\spaceskip=\fontdimen2\font plus
\BIBentryALTinterwordstretchfactor\fontdimen3\font minus
  \fontdimen4\font\relax}
\providecommand{\BIBforeignlanguage}[2]{{%
\expandafter\ifx\csname l@#1\endcsname\relax
\typeout{** WARNING: IEEEtran.bst: No hyphenation pattern has been}%
\typeout{** loaded for the language `#1'. Using the pattern for}%
\typeout{** the default language instead.}%
\else
\language=\csname l@#1\endcsname
\fi
#2}}
\providecommand{\BIBdecl}{\relax}
\BIBdecl

\bibitem{irts2015}
``{The International Technology Roadmap for Semiconductors (ITRS), 2015
  Edition, More Moore},'' http://www.itrs2.net/itrs-reports.html.

\bibitem{engel2011unreliable}
M.~Engel, F.~Schmoll, A.~Heinig, and P.~Marwedel, ``Unreliable yet
  useful--reliability annotations for data in cyber-physical systems,'' in
  \emph{Proc. of WS4C}, 2011.

\bibitem{RDS2001:Jhumka}
A.~Jhumka, M.~Hiller, and N.~Suri, ``Assessing inter-modular error propagation
  in distributed software,'' in \emph{Reliable Distributed Systems, 2001.
  Proceedings. 20th IEEE Symposium on}, 2001, pp. 152--161.

\bibitem{Hiller:2002:PEE:566171.566184}
M.~Hiller, A.~Jhumka, and N.~Suri, ``Propane: An environment for examining the
  propagation of errors in software,'' \emph{SIGSOFT Softw. Eng. Notes},
  vol.~27, no.~4, Jul. 2002.

\bibitem{Meenderinck2009}
C.~Meenderinck, A.~Azevedo, B.~Juurlink, M.~Alvarez~Mesa, and A.~Ramirez,
  ``Parallel scalability of video decoders,'' \emph{Journal of Signal
  Processing Systems}, vol.~57, no.~2, pp. 173--194, 2009.

\end{thebibliography}
\end{document}